\documentstyle[12pt,twoside]{article}
 
\def\a{\alpha}
\def\b{\beta}

\def\d{\delta}
\def\e{\epsilon}                % Also, \varepsilon
\def\f{\phi}                    %       \varphi
\def\g{\gamma}
\def\h{\eta}

\def\l{\lambda}
\def\m{\mu}
\def\n{\nu}
\def\o{\omega}
\def\p{\pi}                     % Also, \varpi
\def\r{\rho}                    %       \varrho
\def\s{\sigma}                  %       \varsigma
\def\t{\tau}

\def\x{\xi}

\def\D{\Delta}

\def\G{\Gamma}

\def\P{\Pi}

\def\cl{{\cal L}}

\catcode`@=11
 
\def\un#1{\relax\ifmmode\@@underline#1\else $\@@underline{\hbox{#1}}$\relax\fi}
 
{\catcode`\'=\active \gdef'{{}~\bgroup\prim@s}}
 
\def\magstep#1{\ifcase#1 \@m\or 1200\or 1440\or 1728\or 2074\or 2488\or
        2986\fi\relax}
 
\def\vpt{\textfont\z@\fivrm
  \scriptfont\z@\fivrm \scriptscriptfont\z@\fivrm
\textfont\@ne\fivmi \scriptfont\@ne\fivmi \scriptscriptfont\@ne\fivmi
\textfont\tw@\fivsy \scriptfont\tw@\fivsy \scriptscriptfont\tw@\fivsy
\textfont\thr@@\tenex \scriptfont\thr@@\tenex \scriptscriptfont\thr@@\tenex
\def\prm{\fam\z@\fivrm}%
\def\unboldmath{\everymath{}\everydisplay{}\@nomath
  \unboldmath\fam\@ne\@boldfalse}\@boldfalse
\def\boldmath{\@subfont\boldmath\unboldmath}%
\def\pit{\@getfont\pit\itfam\@vpt{cmti5}}%
\def\psl{\@subfont\sl\it}%
\def\pbf{\@getfont\pbf\bffam\@vpt{cmbx5}}%
\def\ptt{\@subfont\tt\rm}%
\def\psf{\@subfont\sf\rm}%
\def\psc{\@subfont\sc\rm}%
\def\ly{\fam\lyfam\fivly}\textfont\lyfam\fivly
    \scriptfont\lyfam\fivly \scriptscriptfont\lyfam\fivly
\@setstrut\rm}
 
\def\@vpt{}
 
\def\vipt{\textfont\z@\sixrm
  \scriptfont\z@\sixrm \scriptscriptfont\z@\sixrm
\textfont\@ne\sixmi \scriptfont\@ne\sixmi \scriptscriptfont\@ne\sixmi
\textfont\tw@\sixsy \scriptfont\tw@\sixsy \scriptscriptfont\tw@\sixsy
\textfont\thr@@\tenex \scriptfont\thr@@\tenex \scriptscriptfont\thr@@\tenex
\def\prm{\fam\z@\sixrm}%
\def\unboldmath{\everymath{}\everydisplay{}\@nomath
  \unboldmath\@boldfalse}\@boldfalse
\def\boldmath{\@subfont\boldmath\unboldmath}%
\def\pit{\@subfont\it\rm}%
\def\psl{\@subfont\sl\rm}%
\def\pbf{\@getfont\pbf\bffam\@vipt{cmbx6}}%
\def\ptt{\@subfont\tt\rm}%
\def\psf{\@subfont\sf\rm}%
\def\psc{\@subfont\sc\rm}%
\def\ly{\fam\lyfam\sixly}\textfont\lyfam\sixly
    \scriptfont\lyfam\sixly \scriptscriptfont\lyfam\sixly
\@setstrut\rm}
 
\def\@vipt{}
 
\def\xxxpt{\textfont\z@\thtyrm
  \scriptfont\z@\twfvrm \scriptscriptfont\z@\twtyrm
\textfont\@ne\twfvmi \scriptfont\@ne\twfvmi \scriptscriptfont\@ne\twtymi
\textfont\tw@\twfvsy \scriptfont\tw@\twfvsy \scriptscriptfont\tw@\twtysy
\textfont\thr@@\tenex \scriptfont\thr@@\tenex \scriptscriptfont\thr@@\tenex
\def\unboldmath{\everymath{}\everydisplay{}\@nomath\unboldmath
        \textfont\@ne\twfvmi \textfont\tw@\twfvsy \textfont\lyfam\twfvly
        \@boldfalse}\@boldfalse
\def\boldmath{\@subfont\boldmath\unboldmath}%
\def\prm{\fam\z@\thtyrm}%
\def\pit{\@subfont\it\rm}%
\def\psl{\@subfont\sl\rm}%
\def\pbf{\@getfont\pbf\bffam\@xxxpt{cmbx10\@magscale6}}%
\def\ptt{\@subfont\tt\rm}%
\def\psf{\@subfont\sf\rm}%
\def\psc{\@subfont\sc\rm}%
\def\ly{\fam\lyfam\twfvly}\textfont\lyfam\twfvly
   \scriptfont\lyfam\twfvly \scriptscriptfont\lyfam\twtyly
\@setstrut \rm}
 
\def\@xxxpt{}
 
\def\Huge{\@setsize\Huge{36pt}\xxxpt\@xxxpt}
 
\def\xxxvipt{\textfont\z@\thsirm
  \scriptfont\z@\thtyrm \scriptscriptfont\z@\twfvrm
\textfont\@ne\thtymi \scriptfont\@ne\thtymi \scriptscriptfont\@ne\twfvmi
\textfont\tw@\thtysy \scriptfont\tw@\thtysy \scriptscriptfont\tw@\twfvsy
\textfont\thr@@\tenex \scriptfont\thr@@\tenex \scriptscriptfont\thr@@\tenex
\def\unboldmath{\everymath{}\everydisplay{}\@nomath\unboldmath
        \textfont\@ne\thtymi \textfont\tw@\thtysy \textfont\lyfam\thtyly
        \@boldfalse}\@boldfalse
\def\boldmath{\@subfont\boldmath\unboldmath}%
\def\prm{\fam\z@\thsirm}%
\def\pit{\@subfont\it\rm}%
\def\psl{\@subfont\sl\rm}%
\def\pbf{\@getfont\pbf\bffam\@xxxpt{cmss12\@magscale6}}%
\def\ptt{\@subfont\tt\rm}%
\def\psf{\@subfont\sf\rm}%
\def\psc{\@subfont\sc\rm}%
\def\ly{\fam\lyfam\thtyly}\textfont\lyfam\thtyly
   \scriptfont\lyfam\thtyly \scriptscriptfont\lyfam\twfvly
\@setstrut \rm}
 
\def\@xxxvipt{}
 
\def\HUGE{\@setsize\HUGE{43pt}\xxxvipt\@xxxvipt}
 
\catcode`@=12
 
\font\tenex=cmex10 scaled 1200
\def\Sc#1{\hbox{\sc #1}}        % script "
\def\bo{{\raise.05ex\hbox{\large$\Box$}\:}}             % D'Alembertian
\def\cbo{{\,\raise-.15ex\Sc [\,}}                       % curly "
\def\pa{\partial}                                       % curly d
\def\su{\sum}                                           % summation
\def\TH{{\raise.2ex\hbox{$\displaystyle \bigodot$}\mskip-4.7mu \llap H \;}}
\def\face{\hbox{\normalsize$\;\;\:{\raise.9ex\hbox{\oo n}\mskip-13mu \llap
        {${\buildrel{\hbox{\frtnrm ..}}\over\smile}$}}\:$}}     % happy face
\def\Face{{\raise.2ex\hbox{$\displaystyle \bigodot$}\mskip-2.2mu \llap {$\ddot
        \smile$}}}                                      % another "
\def\Lhat{{\bf\rlap{\kern-.09em$\hat{\phantom L}$}L}}
\def\Lcheck{{\bf\rlap{\kern-.09em$\check{\phantom L}$}L}}
\def\sp#1{{}^{#1}}                              % superscript (unaligned)
\def\sb#1{{}_{#1}}                              % sub"
\def\sl#1{\rlap{\hbox{$\mskip 1 mu /$}}#1}      % good slash for lower case
\def\leftrightarrowfill{$\mathsurround=0pt \mathord\leftarrow \mkern-6mu
        \cleaders\hbox{$\mkern-2mu \mathord- \mkern-2mu$}\hfill
        \mkern-6mu \mathord\rightarrow$}
\def\dvec#1{\vbox{\ialign{##\crcr
        \leftrightarrowfill\crcr\noalign{\kern-1pt\nointerlineskip}
        $\hfil\displaystyle{#1}\hfil$\crcr}}}           % <--> accent
\def\dt#1{{\buildrel {\hbox{\LARGE .}} \over {#1}}}     % dot-over for sp/sb
\def\ddt#1{{\buildrel {\hbox{\LARGE .\kern-2pt.}} \over {#1}}}% double dot-over
\def\frac#1#2{{\textstyle{#1\over\vphantom2\smash{\raise.20ex
        \hbox{$\scriptstyle{#2}$}}}}}                   % fraction
\def\ha{\frac12}                                        % 1/2
\def\sfrac#1#2{{\vphantom1\smash{\lower.5ex\hbox{\small$#1$}}\over
        \vphantom1\smash{\raise.4ex\hbox{\small$#2$}}}} % alternate fraction
\def\bfrac#1#2{{\vphantom1\smash{\lower.5ex\hbox{$#1$}}\over
        \vphantom1\smash{\raise.3ex\hbox{$#2$}}}}       % "
\def\afrac#1#2{{\vphantom1\smash{\lower.5ex\hbox{$#1$}}\over#2}}    % "
\def\boxes#1{
        \newcount\num
        \num=1
        \newdimen\downsy
        \downsy=-1.64ex
        \mskip-7.8mu
        \bo
        \loop
        \ifnum\num<#1
        \llap{\raise\num\downsy\hbox{$\bo$}}
        \advance\num by1
        \repeat}
\def\boxup#1#2{\newcount\numup
        \numup=#1
        \advance\numup by-1
        \newdimen\upsy
        \upsy=.82ex
        \mskip7.8mu
        \raise\numup\upsy\hbox{$#2$}}
 
\newskip\humongous \humongous=0pt plus 1000pt minus 1000pt
\def\caja{\mathsurround=0pt}

\newif\ifdtup
\def\panorama{\global\dtuptrue \openup2\jot \caja
        \everycr{\noalign{\ifdtup \global\dtupfalse
        \vskip-\lineskiplimit \vskip\normallineskiplimit
        \else \penalty\interdisplaylinepenalty \fi}}}
\def\li#1{\panorama \tabskip=\humongous                         % eqalignno
        \halign to\displaywidth{\hfil$\displaystyle{##}$
        \tabskip=0pt&$\displaystyle{{}##}$\hfil
        \tabskip=\humongous&\llap{$##$}\tabskip=0pt
        \crcr#1\crcr}}

\def\NP{Nucl. Phys. B}
\def\PL{Phys. Lett. }

\def\ref#1{$\sp{#1]}$}

\parskip=\medskipamount                 % space between paragraphs (TeX)
\def\baselinestretch{1.2}       % magnification for line spacing (LaTeX)
\thicklines                         % thick straight lines for pictures (LaTeX)
\def\title#1#2#3#4{
\begin{document}
        {\hbox to\hsize{#4 \hfill  #3}}\par
        \begin{center}\vskip.5in minus.1in {\Large\bf #1}\\[.5in minus.2in]{#2}
        \vskip1.4in minus1.2in {\bf ABSTRACT}\\[.1in]\end{center}
        \begin{quotation}\par}
\def\author#1#2{#1\\[.1in]{\it #2}\\[.1in]}

\def\AMIC{Aleksandar Mikovic\'c
\\[.1in]{\it Blackett Laboratory, Imperial College, Prince Consort Road, London
SW7 2BZ, UK}\\[.1in]}

\def\AMICIF{Aleksandar Mikovi\'c\,
\footnote{Work supported by MNTRS and Royal Society}
\\[.1in] {\it Blackett Laboratory, Imperial College, Prince Consort
Road, London SW7 2BZ, UK}\\[.1in]
and \\[.1 in]
{\it Institute of Physics, P.O. Box 57, 11001 Belgrade, Yugoslavia}
\footnote{Permanent address}\\ {\it E-mail:\, mikovic@castor.phy.bg.ac.yu}}

\def\AMSISSA{Aleksandar Mikovi\'c\,
\footnote{E-mail address: mikovic@castor.phy.bg.ac.yu}
\\[.1in] {\it SISSA-International School for Advanced Studies\\
Via Beirut 2-4, Trieste 34100, Italy}\\[.1in]
and \\[.1 in]
{\it Institute of Physics, P.O. Box 57, 11001 Belgrade, Yugoslavia}
\footnote{Permanent address}}

\def\AM{Aleksandar Mikovi\'c 
\footnote{E-mail address: mikovic@castor.phy.bg.ac.yu}
\\[.1in] {\it Institute of Physics, P.O.Box 57, Belgrade 11001, Yugoslavia}
\\[.1in]}

\def\AMsazda{Aleksandar Mikovi\'c 
\footnote{E-mail address: mikovic@castor.phy.bg.ac.yu}
and Branislav Sazdovi\'c \footnote{E-mail: sazdovic@castor.phy.bg.ac.yu}
\\[.1in] {\it Institute of Physics, P.O.Box 57, Belgrade 11001, Yugoslavia}
\\[.1in]}

\def\AMVR{Aleksandar Mikovi\'c\,
\footnote{E-mail address: mikovic@castor.phy.bg.ac.yu}
\\[.1in] 
{\it Institute of Physics, P.O. Box 57, 11001 Belgrade, Yugoslavia}
\\[.2in]
Voja Radovanovi\'c \\[.1 in]
{\it Faculty of Physics, P.O. Box 550, 11001 Belgrade, Yugoslavia}}

\def\endtitle{\par\end{quotation}\vskip3.5in minus2.3in\newpage}
 
% A4
 
\def\endabstract{\par\end{quotation}
        \renewcommand{\baselinestretch}{1.2}\small\normalsize}
 
% Letter
 
\def\xpar{\par}                                         % \par in loops

\def\letterhead{
        \centerline{\large\sf INSTITUTE OF PHYSICS}
        \centerline{\sf P.O.Box 57, 11001 Belgrade, Yugoslavia}
        \rightline{\scriptsize\sf Dr Aleksandar Mikovi\'c}
        \vskip-.07in
        \rightline{\scriptsize\sf Tel: 11 615 575}
        \vskip-.07in
        \rightline{\scriptsize\sf E-mail: MIKOVIC@CASTOR.PHY.BG.AC.YU}}

\def\sig#1{{\leftskip=3.75in\parindent=0in\goodbreak\bigskip{Sincerely yours,}
\nobreak\vskip .7in{#1}\par}}

\def\ssig#1{{\leftskip=3.75in\parindent=0in\goodbreak\bigskip{}
\nobreak\vskip .7in{#1}\par}}

% Referee report
 
\def\ree#1#2#3{
        \hfuzz=35pt\hsize=5.5in\textwidth=5.5in
        \begin{document}
        \ttraggedright
        \par
        \noindent Referee report on Manuscript \##1\\
        Title: #2\\
        Authors: #3}
 
% Book
 
\def\start#1{\pagestyle{myheadings}\begin{document}\thispagestyle{myheadings}
        \setcounter{page}{#1}}
 
% Page and section headings and reference stuff
 
\catcode`@=11
 
\def\ps@myheadings{\def\@oddhead{\hbox{}\footnotesize\bf\rightmark \hfil
        \thepage}\def\@oddfoot{}\def\@evenhead{\footnotesize\bf
        \thepage\hfil\leftmark\hbox{}}\def\@evenfoot{}
        \def\sectionmark##1{}\def\subsectionmark##1{}
        \topmargin=-.35in\headheight=.17in\headsep=.35in}
\def\ps@acidheadings{\def\@oddhead{\hbox{}\rightmark\hbox{}}
        \def\@oddfoot{\rm\hfil\thepage\hfil}
        \def\@evenhead{\hbox{}\leftmark\hbox{}}\let\@evenfoot\@oddfoot
        \def\sectionmark##1{}\def\subsectionmark##1{}
        \topmargin=-.35in\headheight=.17in\headsep=.35in}
 
\catcode`@=12
 
\def\sect#1{\bigskip\medskip\goodbreak\noindent{\large\bf{#1}}\par\nobreak
        \medskip\markright{#1}}
\def\chsc#1#2{\phantom m\vskip.5in\noindent{\LARGE\bf{#1}}\par\vskip.75in
        \noindent{\large\bf{#2}}\par\medskip\markboth{#1}{#2}}
\def\Chsc#1#2#3#4{\phantom m\vskip.5in\noindent\halign{\LARGE\bf##&
        \LARGE\bf##\hfil\cr{#1}&{#2}\cr\noalign{\vskip8pt}&{#3}\cr}\par\vskip
        .75in\noindent{\large\bf{#4}}\par\medskip\markboth{{#1}{#2}{#3}}{#4}}
\def\chap#1{\phantom m\vskip.5in\noindent{\LARGE\bf{#1}}\par\vskip.75in
        \markboth{#1}{#1}}
\def\refs{\bigskip\medskip\goodbreak\noindent{\large\bf{REFERENCES}}\par
        \nobreak\bigskip\markboth{REFERENCES}{REFERENCES}
        \frenchspacing \parskip=0pt \renewcommand{\baselinestretch}{1}\small}
\def\unrefs{\normalsize \nonfrenchspacing \parskip=medskipamount}
\def\Item{\par\hang\textindent}
\def\Itemitem{\par\indent \hangindent2\parindent \textindent}
\def\makelabel#1{\hfil #1}
\def\topic{\par\noindent \hangafter1 \hangindent20pt}
\def\Topic{\par\noindent \hangafter1 \hangindent60pt}

\title{W-Strings on Curved Backgrounds}
{\AMsazda}{3/96}{September 1996}
We discuss a canonical formalism method for
constructing actions describing propagation of W-strings on 
curved backgrounds. The method is based on the construction of a 
representation of 
the W-algebra in terms of currents made from the string coordinates and 
the canonically conjugate momenta.
We construct such a representation for a $W_3$-string propagating in the
background metric with one 
flat direction by using a simple ansatz for the W-generators
where each generator is a polynomial of the canonical currents
and the veilbeins.
In the case of a general background we show that 
the simple polynomial ansatz fails, and terms containing the
veilbein derivatives must be added.

\endtitle

W-string 
theories are higher spin generalizations of ordinary string theories,
such that the string coordinates are not only coupled to 
the world-sheet metric but also to
a set of higher spin world-sheet gauge fields (for a review see [1]). 
Since ordinary string theory can be considered as a gauge 
theory based on the 
Virasoro algebra, one can analogously define a W-string theory as a gauge
theory based on a W-algebra [2] (or any other higher spin conformally 
extended algebra [1]). 
Actions for a large class of W-string theories have been constructed so far
[3-10]. These actions essentially describe a W-string propagating on a flat
background spacetime metric. 
In the case of a curved background metric, the problem of constructing
invariant actions was first considered in [11], where it was solved for a
special case of a group manifold. This construction was based on the
canonical formalism method introduced in ref. [9]. In [11] it was 
crucial that a representation of the W algebra was known in terms of  
the currents 
which obeyed a current algebra associated with the Lie group in question.
However, when the background metric was not a group manifold metric, 
the corresponding canonical
currents did not satisfy a current algebra under the Poisson
brackets. Consequently one could not obtain
a representation for the W algebra in terms of the string coordinates and
the canonicaly conjugate momenta and 
an invariant action could not be constructed. 

In this letter we examine this problem, and show that it is caused by the
ansatz used for constructing the representation of the W generators. 
The ansatz used in [11] is a simple polynomial ansatz where the 
W generators 
are polynomials of the canonical currents and the veilbeins. We show 
that this simple polynomial ansatz also works for an arbitrary 
background metric with one flat direction, while in the most general case
it fails. In conclusions we argue that the simple polynomial ansatz 
must be generalized by adding the terms containing the veilbein
derivatives.  
 
We are going to use the canonical formalism
for constructing the gauge invariant actions [9]. This method works if one
knows a representation of the algebra of gauge symmetries in terms of the
coordinates and canonically conjugate momenta. The basic idea is simple:
given a set of canonical pairs $(p\sb i  , q\sp i )$ 
together with the Hamiltonian $H\sb 0 (p,q)$ and the constraints
$G\sb \a (p,q)$ such that
$$ \{G\sb \a, G\sb \b \} = f\sb{\a\b}\sp \g G\sb \g \quad,\eqno(1)$$
$$ \{G\sb \a, H\sb 0 \} = h\sb{\a}\sp \b G\sb \b \quad,\eqno(2)$$
where $\{,\}$ is the Poisson bracket and (1) is the desired algebra
of gauge symmetries, then
the corresponding action is given by
$$ S = \int dt \left( p\sb i \dt{q}\sp i - H\sb 0 - \l\sp \a G\sb \a 
\right) \quad.\eqno(3)$$
The parameter $t$ is the time and dot denotes time derivative. 
The Lagrange multipliers $\l\sp \a (t)$ play the role of the
gauge fields associated with the gauge symmetries generated by $G_\a$.
The indices $i,\a$ can take
the discrete as well as the continious values.
Note that the coefficients $f\sb{\a\b}\sp{\g}$ and $h\sb{\a}\sp{\b}$ can
be arbitrary functions of $p_i$ and $q^i$, and hence the algebra (1) is 
general
enough to accomodate the case of the $W$ algebras, where the right-hand
side of the Eq. (1) is a non-linear function of the generators.
The action $S$ is invariant under the following gauge transformations
$$ \li{ \d p\sb i = &\e\sp \a \{ G\sb \a , p\sb i \} \cr
        \d q\sp i = &\e\sp \a \{ G\sb \a , q\sp i \} \cr
        \d \l\sp \a = &\dt{\e}\sp \a - \l\sp \b \e\sp \g f\sb{\g\b}\sp \a 
                      - \e\sp \b h\sb \b\sp \a \quad.&(4)\cr}$$
It can be seen from the the transformation law for
$\l\sp \a$ why they can be identified as gauge fields.

Since we want to describe propagation of a bosonic W-string on a curved
background, the canonical coordinates will be a set of two-dimensional (2d) 
scalar fields $\f\sp A (\s,\t)$, $A=1,...,N$, where $\s$ is the string 
coordinate ($0\le \s \le 2\p$) and $\t$ is the evolution parameter. 
$\f^A$ will be coordinates on an $N$-dimensional space-time manifold $M$.
On $M$ is also given a metric $G_{AB}$, which can be of arbitrary 
signature.
Let $\p\sb A (\s,\t)$ be the canonically conjugate momenta, satisfying
$$ \{ \f\sp A (\s_1 ,\t), \p\sb B (\s_2 ,\t)\} = 
\d^{A}_{B}\d (\s_1  - \s_2 ) \quad.
\eqno(5)$$
In order to construct an invariant action, we will need a canonical 
representation of the corresponding W-algebra. 
We start from the action for an ordinary bosonic string propagating on $M$,
which is a 2d $\s$-model action
$$
S_2=\int d^2 \s {1\over 2}\sqrt{-g}g^{\m \n}\pa_\m \f^A \pa_\n \f^B G_{AB}(\f)
\quad. \eqno(6)
$$
The action (6) can be rewritten in a canonical form as
$$ S\sb 2 = \int_{\t_1}^{\t_2} d\t \int_{0}^{2\p}d\s \left( \p\sb A 
\dt{\f}\sp A - h\sp{\a} T\sb{\a}  \right)\quad,\eqno(7)$$
where $\a=\{+,-\}$ and the constraints $T_\a$ are given by
$$ T_\pm ={1\over 2} G^{AB} J_{\pm A} J_{\pm B}
\quad.\eqno(8) $$ 
We have introduced the currents
$$J_{\pm A}={1\over \sqrt2} (\p_A \pm G_{AB} \f^{\prime B} ) 
\quad,\eqno(9)$$
and the primes stand for the $\s$ derivatives. The constraints $T_\a$ are 
$++$ and $--$ components of the string energy-momentum tensor $T_{\m\n}$, 
where 
$x^\pm = x^0 \pm x^1$. $T_\a$ satisfy the classical Virasoro algebra 
$$ \{T\sb \pm (\s_1),T\sb \pm (\s_2)\} 
= \pm \d^{\prime} (\s_1 - \s_2 ) (T\sb \pm (\s_1) +
T\sb \pm (\s_2)) \eqno(10)$$
under the Poisson brackets (5).
The Poisson brackets of the currents $J_{\a A}$ are given by
$$ \li{ \{J\sb{\pm A} (\s_1),J\sb{\pm B} (\s_2)\} 
=& \pm {1\over 2}(\pa_B G_{AC}- \pa_A G_{BC}) \f^{\prime C} 
\d(\s_1-\s_2)\cr
&\pm {1\over 2} [G_{AB}(\s_1)+G_{AB}(\s_2)] \d^{\prime} (\s_1 - \s_2 )
                  \quad.&(11)\cr}$$
For a general metric $G_{AB}$ the relations (11)
do not form a current algebra, and the currents of opposite chirality 
do not have vanishing Possion brackets, but 
$$\li{
\{J_{+A} (\s_1), J_{-B}(\s_2) \}=&\f^{\prime C} \G_{C,AB} \d(\s_1-\s_2)\cr
 \equiv & 
\f^{\prime C} {1\over 2}(\pa_A G_{BC} +\pa_B G_{AC}-\pa_C G_{AB}) 
\d(\s_1-\s_2),  &(12)\cr}
$$
where $\G_{C,AB}$ is the Christoffel symbol. When $G_{AB}$ is a group 
manifold metric, then the relations (11) and (12) can give two independent 
chiral current algebras [11].

Although the currents (9) do not form a closed current algebra, we will 
proceed with
the ansatz from the group metric case in order to find its limitations.
We take
$$
W\sb{\a s} = {1\over s} D\sp{A\sb 1 \cdots A\sb s}J\sb{\a A\sb 1}
\cdots J\sb{\a A\sb s} \quad (s=2,...,N)\quad, \eqno(13)
$$
where the coefficients $D\sp{A_1 ... A_s}$ will be determined from the
requirement that the Poisson brackets of the quantities (13) form a
$W$ algebra. Note that from (8) we have $D^{AB} = G^{AB}$.
For the sake of simplicity we specialize to the
$W_3$-string case. The expressions (13) should then obey a
classical $W_3$ algebra ($W\equiv W_3$) 
$$\li{\{T\sb \pm (\s_1),T\sb \pm (\s_2)\} 
&= \pm\d^{\prime} (\s_1 - \s_2) (T\sb \pm (\s_1) +
T\sb \pm (\s_2)) &   (14a)  \cr
\{T\sb \pm (\s_1),W\sb \pm (\s_2)\} 
&= \pm\d^{\prime} (\s_1 - \s_2) (W\sb \pm (\s_1) +
2W\sb \pm (\s_2)) & (14b) \cr
\{W\sb \pm (\s_1),W\sb \pm (\s_2)\} 
&= \pm \d^{\prime} (\s_1 - \s_2) (T\sp 2\sb \pm (\s_1) +
T^2\sb \pm (\s_2)) \quad.&(14c)\cr}$$
From the relation (14b) it follows that $D^{ABC}$ is a covariantly 
constant tensor
$$
\nabla_D D^{ABC} = \pa_D D^{ABC} + \G^{(A}_{DE} D^{BC)E} =0 \quad, 
\eqno(15)
$$
while from the relation (14c) it follows that  
$$
D^{(AB}{}_E D^{CD)E} ={1\over 2} G^{(AB} G^{CD)} \quad.    \eqno(16) 
$$
If we introduce the veilbeins $E_a^A(\f)$ as
$$
G^{AB} = \h^{ab} E_a^A E_b^B  \quad,       \eqno(17)
$$
where $\h_{ab}$ is a flat metric,
then the equation (15) is satisfied if
$$
D^{ABC}=\D^{abc} E_a^A E_b^B E_c^C   \quad,     \eqno(18)
$$
where $\D^{abc}$ are $\f$-independent coefficients and 
$$ \nabla_A E_b^B= \pa_A E_b^B + \G_{AC}^B E_b^C = 0 \quad.\eqno(19)$$
Then the condition (16) becomes 
$$
\D^{(ab}_e \D^{cd)e} ={1\over 2} \d^{(ab} \d^{cd)}\quad,  \eqno(20)
$$
which is satisfied for $\D^{abc} = d^{abc}$, where $d^{abc}$
are the flat-background coefficients [9]. However, there are further 
conditions which $D^{ABC}$ have to satisfy, and they come from the 
requiriments
$$ \{T_+ , W_- \} =0 \quad,\quad\{T_- , W_+ \} =0 \quad,
\quad \{W_+ , W_- \} = 0 \quad.\eqno(21)$$
Equations (21) give the following constraints
$$\li{0 =& -\frac13 G^{AB}\pa_B D^{CDE} J_{\pm A} J_{\mp C} +
           \frac12 \pa_C G^{AB}D^{CDE} J_{\pm A} J_{\pm B}
   \pm\sqrt{2}  G^{AB} D^{CDE}\G_{F,AC} J_{\pm B} \f^{\prime F} \cr
      0=&  -\frac13 D^{ABC}\pa_C D^{DEF} J_{+ A} J_{+ B} J_{- D} 
             +\frac13 \pa_D D^{ABC} D^{DEF} J_{+A} J_{+B} J_{+C} \cr 
+ &\sqrt{2} D^{ABC} D^{DEF}\G_{H,CD} J_{+ A} J_{+ B} \f^{\prime H}
\, . &(22)\cr}$$
By using (9), the constraints (22) can be rewritten as polynomials in 
$\p_A$ and $\f^{\prime A}$, and (22) will be satisfied if the coeficients
of these polynomials vanish. In this way one obtains the additional
constraints on the solution (18). Hence it is clear that (18) is a solution
only for a special class of backgrounds, and one example is the 
group manifold case [11]. A second example will be constructed here, and
it is given by a background metric with one flat direction. 
Note that if one
wants to find new solutions, the strategy of solving the constraints (15)
(16) and (22) is not very efficient, and better thing to do is to
look among existing representations of $W_3$ in terms of scalar fields.

In order to construct the solution for the background with one flat
direction, we start from a flat-background representation 
$$T_{\pm}=\ha \h^{ab} J_{\pm a} J_{\pm b} \quad,
\quad W = \frac13 d^{abc} J_{\pm a} J_{\pm b} J_{\pm c}\quad, \eqno(23)$$
where the non-zero $d^{abc}$ are given by
$$ d^{111}={1\over\sqrt{2}} \quad,\quad d^{1ij} = -{\h^{ij}\over\sqrt{2}}
\quad,\quad (i=2,...,N) \quad.\eqno(24)$$
Equation (23) can be then rewritten as
$$ T_{\pm} = \ha\h^{11} J_{\pm 1}^2 + T_{\pm 2} \quad,
\quad W_\pm ={1\over3\sqrt{2}} J_{\pm 1}^3 -\sqrt{2} J_{\pm 1} T_{\pm 2}
\quad,\eqno(25)$$
where $T_{\pm 2} = \ha \h^{ij} J_{\pm i} J_{\pm j}$ is the 
energy-momentum tensor of the
fields $\f^i$. Note that the representation (25) is valid even 
when $T_2$ is an arbitrary 2d energy-momentum tensor as long as
$\{J_1,T_2\} =0$. This property allows us to construct a $W_3$ 
representation for a curved background with one flat direction 
$$G_{11}=\pm 1\quad,\quad G_{1i} =0\quad,\quad G_{ij}= g_{ij} \quad,
\quad\pa_1 g_{ij} =0\quad,\eqno(26)$$
where $g_{ij}$ is an arbitrary $(N-1)$-dimensional metric. Namely, $T_2$
in that case follows from the formulas (8) and (9), while $J_1$ is given
by (9). The corresponding representation is given by the formula (25),
and it is of the form (18), where now $E_a^A$ are associated to the 
metric (26). 

The $W_3$-string action in the background (26) now follows from the general
formula (3)
$$ S\sb 3 = \int_{\t_1}^{\t_2} d\t \int_0^{2\p} d\s 
\left( \p\sb A \dt{\f}\sp A -   
\su_{s=2}^3 b\sb s\sp{\a} W\sb{\a s}\right) \quad,\eqno(27)$$
where 
$$ T_{\pm} = \ha G^{11} J_{\pm 1}^2 + \ha g^{ij} J_{\pm i} J_{\pm j} \quad,
\quad W_\pm ={1\over3\sqrt{2}} J_{\pm 1}^3 
-{1\over\sqrt{2}}g^{ij} J_{\pm 1}  J_{\pm i} J_{\pm j}
\quad.\eqno(28)$$
As discussed in [9] the 2d diffeomorphism invariance requires
$H\sb 0 =0$, while $b_s^\a$ are the
lagrange multipliers, which are also the gauge fields corresponding to
the $W$-symmetries. The gauge transformation laws can be determined from
the Eq. (4), and in the $W\sb 3$ case  we obtain
$$\li{\d \p\sb A = & {1\over 2} \{ \e^\a \pa_A G^{BC}J_{\a B}
J_{-\a C}  + \x^\a D^{BCD}[ \pa_A G_{BE} J_{-\a}^E + (\pa_B
G_{EA}-\pa_EG_{BA})J_{\a}^E ] J_{\a C}J_{\a D}\} \cr
-&{(-)^\a \over \sqrt 2 } \left( \e^\a J_{\a A} +\x^\a D\sb A\sp{BC} 
J\sb{\a B} J\sb{\a C} \right)^{\prime}  ,&(29.a) \cr
\d \f\sp A =& -{1\over \sqrt 2}(\e\sp \a  J\sb{\a}\sp{A}  
+\x\sp \a D\sp{ABC} J\sb{\a B} J\sb{\a C})  
\quad,&(29.b)\cr
\d h\sp \a =& \dt{\e}\sp \a -(-1)^\a [ h\sp \a (\e\sp \a)^{\prime} -
(h\sp \a )^{\prime}\e\sp \a] + (-1)^\a [\x\sp \a (b\sp \a)^{\prime} -
(\x\sp \a)^{\prime} b\sp \a ] T\sb \a \quad,&(29.c)\cr
\d b\sp \a =& \dt{\x}\sp \a +(-1)^\a \left[ 2(h\sp \a)^{\prime}\x\sp \a
- h\sp \a (\x\sp \a )^{\prime} - 2 b\sp \a (\e\sp \a)^{\prime}
+ (b\sp \a )^{\prime}\e\sp \a\right] \quad,&(29.d)\cr}$$
where $h^\a = b_2^\a$, 
$b^\a = b_3^\a$, $\e^\pm$ are the parameters of
the $T\sb \pm$ transformations, while $\x^\pm$ are the parameters of the
$W\sb \pm$ transformations.
In all equations we use Einstein's summation convention, i.e. summation is
performed only if the up and down index are the same. 

In order to find a geometrical interpretation of the action (27) we need
to know its second order form.
It can be obtained by replacing the momenta $\p\sb A$ in (27) by their 
expressions in terms of $\f^A$. These expressions can be obtained
from the equation of motion 
$${\d S_3 \over \d \p\sb A} = \dt{\f}\sp A - {h\sp \a \over \sqrt 2} 
J_\a^A -{1 \over \sqrt 2} b\sp \a D\sp{ABC} J_{\a B} J_{\a C} = 0 
\quad.\eqno(30)$$
This is a quadratic equation in  $\p_A$, and therefore the second order 
form of the Lagrangean density of (27)
will be a non-polynomial function of $\pa_\m \f^A$, 
$h^\a$ and $b^\a$. Since every solution of (30) can be written
as an infinite power series in $\pa_\m \f^A$, the Lagrangean density
will also be an infinite power series in $\pa_\m \f^A$, in a complete 
analogy with the flat background case [9] and the group manifold case
[11].

In the $W_2$ case one can show that
after the elimination of the momenta in (7)
one obtains the covariant
action (6), after the following identifications 
$$ \tilde{g}\sp{00} = {2\over h\sp + + h\sp -} \quad,\quad
\tilde{g}\sp{01} = {h\sp - - h\sp + \over h\sp + + h\sp -} \quad,\quad
\tilde{g}\sp{11} = - {2h\sp + h\sp -\over h\sp + + h\sp -} \quad,
\eqno(31)$$
where $\tilde{g}\sp{\m\n} = \sqrt{-g}g\sp{\m\n}$. The covariant form of the
2d diffeomorphism transformations can be obtained from the Eq. (29.b), 
by rewritting it as
$$ \d\f\sp A =- {\e\sp \a \over \sqrt 2} J\sb{\a}\sp{A} =
-{\e\sp \a\over\sqrt{ h^+ + h^- }}\tilde{e}\sb \a\sp \m \pa\sb \m 
\f\sp a = \e\sp \m \pa\sb \m \f\sp a \quad,\eqno(32)$$
where
$$ \tilde{e}\sb{\a}\sp \m = {1\over \sqrt{h\sp + + h\sp -}}
\pmatrix{1 &h\sp - \cr 1 &-h\sp +\cr} \quad.\eqno(33)$$
Eq. (29.c) can be rewritten as
$$\d\tilde{g}\sp{\m\n} = -\pa\sb \r (\e\sp \r \tilde{g}\sp{\m\n}) +
\pa\sb \r \e\sp{(\m|}\tilde{g}\sp{|\n)\r} \quad,\eqno(34)$$
which is the
diffeomorphism transformation of a densitized metric
generated by the parameter $\e\sp \m$. The metric $g\sp{\m\n}$ can be 
written as
$$g\sp{\m\n}= {1\over \sqrt{-g}(h\sp + + h\sp - )}
\pmatrix{2 &h\sp - - h\sp +\cr h\sp - - h\sp + &-2h\sp + h\sp - \cr}
= e\sb +\sp{(\m|} e\sb -\sp{|\n)} \quad, \eqno(35)$$
where $e\sb \a\sp \m =(-g)^{-{1 \over 4}} \tilde{e}\sb \a\sp \m$ are 
the zweibeins. 
Note that $\sqrt{-g}$ remains undetermined, because
the action (6) is independent of $\sqrt{-g}$ due to the Weyl 
symmetry
$$ \d g\sp{\m\n} = \o g\sp{\m\n}\quad. \eqno(36) $$
Also note that the relations  (31),(33) and (35) are essentialy the same 
as in the flat background case [9], as well as in the group manifold case 
[11].

In the $W\sb 3$ case we have from the Eq. (30)
$$\li{ \p_A  &= 
G_{AB}\,\tilde{g}^{0\m}\pa_\m \f^B + \sqrt{2 \tilde{g}^{00}} \d\P_A \cr
\d\P_A &= -{\sqrt {\tilde{g}^{00}} b\sp \a \over 2}D\sb{A}\sp{BC}J_{\a B}
J_{\a C} \equiv -{1 \over 2}B^\a D_A{}^{BC} {\bar J}_{\a B}
{\bar J}_{\a C} \quad, &(37)\cr}$$
where 
$$ {\bar J}_\a^A =(\tilde{g}^{00})^{-\ha} J_\a^A \quad,\quad 
B^\a = (\tilde{g}^{00})^{\frac32} b^\a \quad,\eqno(38)$$
and $\tilde{g}^{00}$ is given by the Eq. (31).
Then the action (27) takes the following form
$$ S\sb 3 = \int d^2 \s \left( \cl_{2} -\d\P^A \d\P_A
-{ B\sp{\a}\over 3}D\sp{ABC}{\bar J}_{\a A}{\bar J}_{\a B}{\bar J}_{\a C} 
\right)
\quad,\eqno(39)$$
where $\cl_2$ is the Lagrangean density of the action (6). Note that the
action (39) can be written in a more elegant form
$$ S\sb 3 = \int d^2 \s \left( {\bar J}_{+A}\pa_-\f^A+
{\bar J}_{-A}\pa_+\f^A -{\bar J}_+^A {\bar J}_{-A}
-{ B\sp{\a}\over 3}D\sp{ABC}{\bar J}_{\a A}{\bar J}_{\a B}{\bar J}_{\a C} 
\right)
\quad,\eqno(40)$$
where we have used the equation (30) rewritten as
$$ {\bar J}\sb{\pm}^A =\pa\sb \pm \f\sp A - 
\ha B\sp \a D\sp{ABC}{\bar J}_{\a B} {\bar J}_{\a C}
\quad.\eqno(41)$$
Eq. (41) can be used to obtain a power series expansion of ${\bar J}_A$ 
in terms of $\pa\sb \pm\f$, $h$ and $B$, which can be inserted into Eq. 
(40) to give 
the  corresponding power series expansion of the action. Up to the second 
order in $B$ the Lagrangean desity can be written as
$$\li{ \cl_3 = &\cl_{2} -\frac13 B\sp{\a} D\sb{ABC}\pa\sb{\a}\f\sp A
\pa\sb{\a}\f\sp B \pa\sb \a\f\sp C\cr  +& 
\frac14 B\sp{\a}B\sp{\b}D\sb{ABC}D\sp A\sb{DE}\pa\sb{\a}\f\sp B
\pa\sb{\a}\f\sp C  \pa\sb{\b}\f\sp D \pa\sb{\b}\f\sp E +
O(B^3)\quad.&(42)\cr}$$

It is clear that the covariant form of (42) will be
$$\li{\cl_3 =& \ha\tilde{g}^{\m\n}G^{AB}(\f)
\pa\sb \m \f\sp A \pa\sb \n \f\sp B +  
\frac13\tilde{b}\sp{\m\n\r}D\sb{ABC}(\f)
\pa\sb \m \f\sp A \pa\sb \n \f\sp B \pa\sb \r \f\sp C \cr
&+ \frac14\tilde{c}\sp{\m\n\r\s}D\sb{AB}\sp{E}(\f)D\sb{ECD}(\f)
\pa\sb \m \f\sp A \pa\sb \n \f\sp B \pa\sb \r \f\sp C \pa\sb \s \f\sp D
+ \cdots \quad,&(43)\cr}$$
where the objects $\tilde{g}$, $\tilde{b}$, $\tilde{c}$,
... , must transform
as 2d tensor densities since the action is invariant under the 
infinitesimal diffeomorphisms
$$ \d \f^A = -{\e^A\over{\sqrt 2}} J_\a^A = {\e}^\m \pa_\m \f^A \quad,
\quad {\e}^\m = f^\m (\e^\pm , h^\pm ,
b^\pm,\f^A, \pa_\m \f^A ) \quad.\eqno(44)$$
The action (43) is also invariant under a
generalized Weyl symmetry [1]. This symmetry is built in
by our construction, since we used only four
independent gauge fields $h\sp \pm$ and $B\sp \pm$. Consequently, 
the fields $\tilde{g}$,
$\tilde{b}$, $\tilde{c}$, ... are functions of $h$ and $b$, 
and one can check order by order in $\pa\f$ that
$$ \tilde{g}\sb{\m\n}\tilde{b}\sp{\m\n\r} = 0 \quad,\quad
   \tilde{c}\sp{\m\n\r\s} = \tilde{g}\sb{\t\e}\tilde{b}\sp{\m\n\t}
\tilde{b}\sp{\e\r\s} \quad, \eqno(45)$$
and so on. Therefore the independent gauge fields are the 2d metric 
$g_{\m\n}$ and a symmetric tensor $b_{\m\n\r}$, related to ${\tilde g}$
and ${\tilde b}$ as
$$ {\tilde g}_{\m\n} = {\sqrt {-g}}g_{\m\n} \quad,\quad 
{\tilde b}_{\m\n\r} = {\sqrt {-g}}\left( b_{\m\n\r} - \frac32 g_{(\m\n|}
b\sp{\s}\sb{\s|\r)}\right)\quad. \eqno(46)$$

In conclusion we can say that the second order Lagrangean (43) for the 
$W_3$-string
propagating in the background metric (26)
has the same form as in the group-manifold case [11]. The appearance
of the same generic form for the action in both cases may lead one to 
belive that (43) is valid for a general case. However, it is clear from
our analysis that the relation (18) cannnot be satisfied for a general
background metric. In that case one can try to generalize
Eq. (18) by allowing $\D^{abc}$
to become functions of $\f^A$. Since $\D^{abc}$ must be built out 
of $\h_{ab}$, $d^{abc}$ and the veilbeins, the only way to make $\D^{abc}$
$\f$-dependent is to allow the veilbein derivatives to appear in the 
expressions for $\D^{abc}$. For example 
$$\D^{abc} = d^{abc} + c_1\pa_A E^{Aa}\pa_B E^{Bb}\pa_C E^{Cc} + ... 
\quad,\eqno(47)$$
where $c_1$ is a numerical constant.
Given the fact that $D^{ABC}$ has to be highly constrained, it is 
difficult to see whether the modification (47) can yield a solution in
terms of the unconstrained vielbeins. A more likely possibility is to
change the original ansatz for the W generators (13), by adding 
the terms in $W_s$ which 
contain higher powers of $J$ then the conformal spin $s$.
The coefficients
of these terms must be proportional to the vielbein derivatives,
since they vanish in the flat case. A more precise functional
dependence can be obtained
from the fact that these terms also vanish in the group-manifold case 
and in the case of the metric (26). In the lowest order, the new ansatz
can be written as
$$ W_{\a 3} = \frac13 D^{ABC} J_{\a A} J_{\a B} J_{\a C} + 
F_\a^{ABCD} J_{\a A} J_{\a B} J_{\a C} J_{\a D} 
              + ... \quad,\eqno(48)$$ 
where $F_\a$ is built from $\h_{ab}, d_{abc}$, veilbeins and 
the derivatives of veilbeins. 
For example $\nabla^{(A} D^{BCD)}$ would be a consistent term in $F^{ABCD}$. 
However, an exact determination of $W$ would require a separate paper, and
it would be interesting to find out whether the power series in $J$
in (48) terminates
after a finite number of terms and whether the terms containing the opposite 
chirality currents have to be added.

Acknowledgements: We would like to thank MNTRS for a financial
support and Simon Ljakhevich for usefull
discussions.

\refs

\Item{[1]}C.M. Hull, Classical and quantum W-gravity, in Strings and
Symmetries 1991, eds. N. Berkovits et al., World Scientific, Singapore 
(1992)
\Item{[2]}A.B. Zamolodchikov, Teor. Mat. Fiz. 65 (1985) 1205\\
          V.A. Fateev and A.B. Zamolodchikov, \NP 280 (1987) 644\\
          V.A. Fateev and S. Lukyanov, Int. J. Mod. Phys. A3 (1988) 507
\Item{[3]}C.M. Hull, \PL 240B (1990) 110
\Item{[4]}K. Schoutens, A. Sevrin and P. van Nieuwenhuizen, \PL 243B (1990) 245
\Item{[5]}E. Bergshoeff, C.N. Pope, L.J. Romans, E. Sezgin, X. Shen and 
          K.S. Stelle, \PL 243B (1990) 350
\Item{[6]}C.M. Hull, \NP 353 (1991) 707
\Item{[7]}C.M. Hull, \PL 259B (1991) 68
\Item{[8]}A. Mikovi\'c, \PL 260B (1991) 75
\Item{[9]}A. Mikovi\'c, \PL 278B (1991) 51
\Item{[10]}G.M. Sotkov, M. Stanishkov and C.-J. Zhu, \NP 356 (1991) 245, 
439
\Item{[11]}A. Mikovi\'c and B. Sazdovi\'c, Mod. Phys. Lett. A 10
 (1995) 1041

\end{document}